\documentclass[times, 10pt,twocolumn]{article} 
\usepackage{latex8}
\usepackage{times}
\usepackage{url}

\usepackage[dvips]{graphics}
\usepackage[dvips]{graphicx}
\usepackage[dvips]{epsfig}
\usepackage{graphicx}
\usepackage{amssymb}

\makeatletter
\def\subsubsection{\@startsection {subsubsection}{3}{\z@}
   {10pt plus 2pt minus 2pt}{10pt plus 2pt minus 2pt} {\elvbf}}

\makeatother

\begin{document}

\title{Revealing subnetwork roles using contextual visualization: comparison of metabolic networks}

\author{Romain Bourqui  \\
LaBRI, Universit\'e Bordeaux I, France \\
romain.bourqui@labri.fr \\
\and
Fabien Jourdan\\
INRA, UMR1089, X\'enobiotiques\\
F-31000 Toulouse, France\\
Fabien.Jourdan@toulouse.inra.fr}

\maketitle
\thispagestyle{empty}

\begin{abstract}
This article is addressing a recurrent problem in biology: mining newly built large scale networks.
Our approach consists in comparing these new networks to well known ones.
The visual backbone of this comparative analysis is provided by a network classification hierarchy.
This method makes sense when dealing with metabolic networks since comparison could be done using pathways (clusters).
Moreover each network models an organism and it exists organism classification such as taxonomies.\\
Video demonstration:\\
\small\url{http://www.labri.fr/perso/bourqui/video.wmv}

\end{abstract}
	
\section{Background and motivation}
Visual mining of large networks is a challenging problem in biology since more and more large networks are inferred from high-throughput experiments (protein-protein interaction networks, metabolic networks \cite{biocyc} and gene networks \cite{Jourdan2008}).
The challenge is to understand the biological functions of their different parts.
A way to circumvent this problem consists in fitting parts of the data onto available knowledge.
For instance when discovering a new biological network, if some elements had already been assigned to a given a function then they will probably behave in a similar way in the new network.

\begin{figure}[ht]
	\centering
		\includegraphics[width=.9\linewidth]{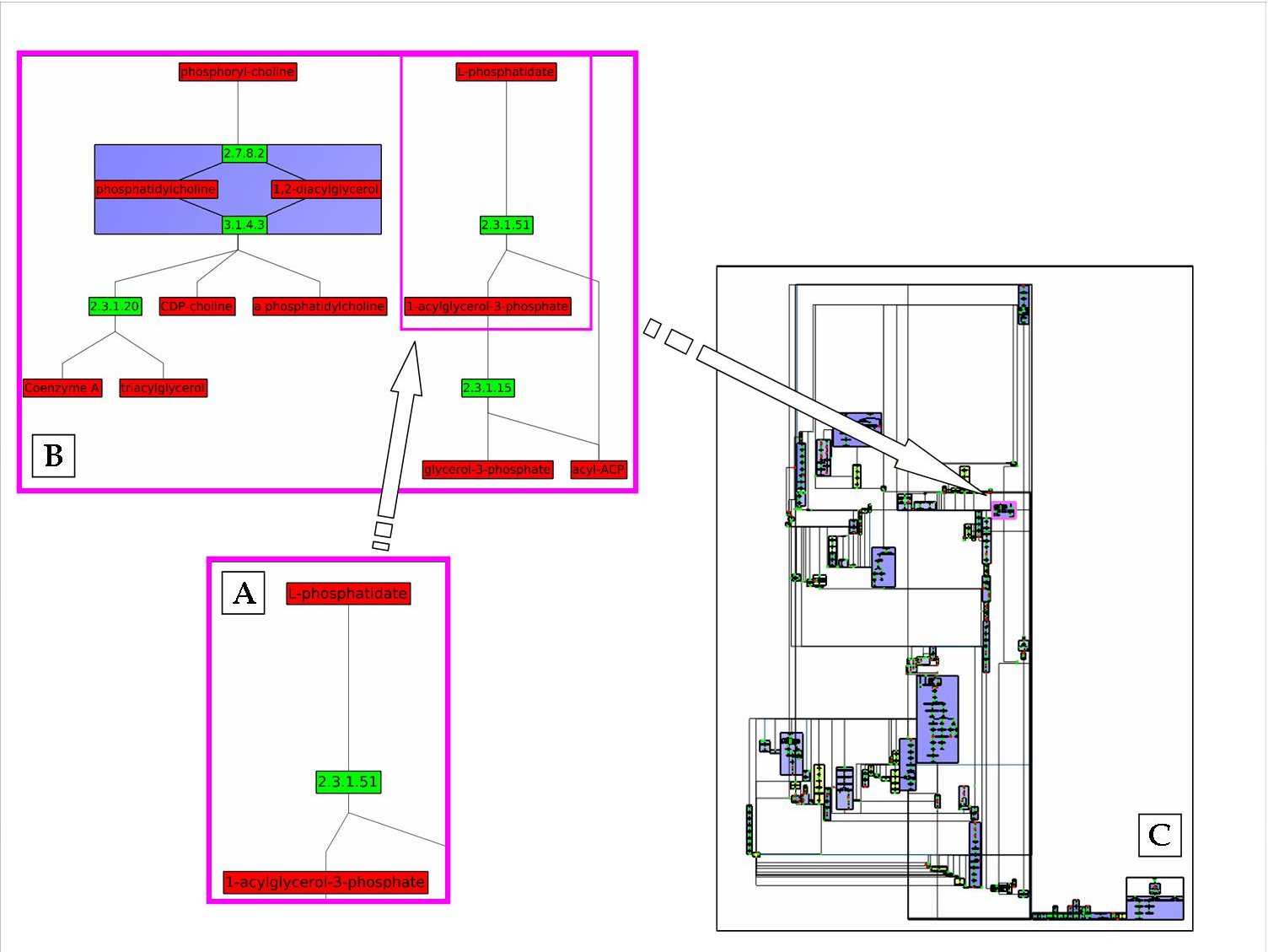}
	\caption{The different scales of metabolic modeling. First scale a metabolic reaction turn a metabolite (in red) into another one under the action of an enzyme (in green)(A). A set of reaction will correspond to a metabolic pathway (B) which is a subgraph of the entire metabolic network (C).}
	\label{fig:metabolism}
\end{figure}
Our collaboration with biologists led us to focus on a particular biological research topic: metabolism. 
Metabolism is the set of biochemical reactions (figure \ref{fig:metabolism}A) that are used to perform vital biological functions such as energy generation.
Each metabolic function is modelled by a set of interconnected reactions corresponding to a small graph called a metabolic pathway (figure \ref{fig:metabolism}B) \cite{VWGW02}.
Since the output of a pathway is often the input of another pathway it is possible to merge all these pathways into a single metabolic network (figure \ref{fig:metabolism}C).
Each organism has its own metabolic network.
For instance mammalians and plants won't have the same metabolic network since only plants can generate energy using the photosynthesis pathway.
But on the other hand they will share biological functions, that are pathways.
When biologists are discovering newly inferred metabolic networks, they have to make this kind of comparison.
But they are dealing with networks containing hundreds of elements.
Thus, the challenge is to provide a visualization tool allowing them to easily mine new metabolic networks by comparing them to already known ones. 
Based on their observations they will address the following questions: which metabolic functions are shared by these organisms? Is it possible to find a metabolic core between different organisms?

The next section will describe the task defined in collaboration with biologists and the related visualization challenges.
Then we will present the data and model used to build the visualization that will be described in the last section.

\section{Task and challenges}\label{task}
Comparative study of biological networks is a powerful approach in system biology since it uses available knowledge to interpret new networks.
A first way to compare networks consists in looking for topologically similar subnetworks.
This approach is well suited to understand the evolution of organisms.
In fact two topologically similar parts generally come from a duplication in the genome during the evolution.
The comparison of networks is a computationally difficult problem (see the graph isomorphism problem in \cite{GJ79}).
Nevertheless heuristics had been proposed to align metabolic pathways \cite{Pinter05}.
But the issue is then to scale to the size of metabolic networks since they are tenth time larger than pathways.
To overcome this problem we propose to use the annotations of these networks when they are available; for instance by using labelled nodes or group of nodes (clusters).
It is then easier to identify common subparts since the number of cluster is much lower than the number of nodes (e.g. boxes on figure \ref{fig:metabolism}C).

In this article we will focus on a particular study case which raises two more generic questions: comparing clustered networks and comparing a set of networks with already known ones.
In particular we will focus on a set of organisms called \emph{Alphaproteobacteria}.
It is a sub-group of \emph{Proteobacteria} which are a major group (phylum) of bacteria.
They include a wide variety of pathogens, such as Escherichia, Salmonella, Vibrio, Helicobacter, and many other notable genera.
It exist different kinds of \emph{Alphaproteobacteria}, in particular we will focus on three of them: \emph{Rickettsiales} (pathogen which causes a variety of diseases in humans), \emph{Caulobacter vibroides} (a bacterium essential for the carbon cycle) and \emph{Agrobacterium tumefaciens} (bacterium responsible for tumors in plants).
These genomes had been recently sequenced and consequently new metabolic networks were built.
Based on this data, our first aim is to help biologists in their understanding of the different metabolic properties of each \emph{Alphaproteobacteria}.
Moreover this analysis will be enhanced by adding a context to this comparison.
The context will be provided by supplementary knowledge: metabolic networks of other organisms (for instance other \emph{Proteobacteria}).

A challenge raised by the biological questions that our visualization is addressing is the integration of different representation scales: pathway, network, organism.
Thus it is necessary to draw: metabolic pathways, metabolic networks and a backbone structure connecting them.
To draw metabolic pathways and networks we are going to use dedicated graph drawing algorithms.
The challenge is then to embed these drawings in the representation space.
Since we are dealing with a comparative task we need a structure that highlight a logical organization of organisms.
A particularly well suited structure is the hierarchy since it allows abstraction.
Indeed in a hierarchical classification each internal node models the common information contained in all underneath nodes.
In our study case each internal node contains shared pathways.

In biology it exists several ways to build a hierarchy: taxonomies (trees), phylogenies (trees) or ontologies (directed acyclic graph).
For the \emph{Alphaproteobacteria} task we chose to use the taxonomy.
But it is important to note that our approach is generic enough to allow the use of any other kind of hierarchy.

Finally the main challenge is due to the fact that the data we propose to visualize is quite complex since it contains metabolic networks of $29$ different 
organisms. These networks are made of $21552$ vertices and $27565$ edges composed of $4541$ pathways. 
These pathways share $629$ different names	.
This large dataset also raises a navigation problem since it is not possible to visualize all the information details simultaneously.

\subsection{Pathway oriented comparison}
Our method relies on the notion of metabolic pathways which provides a clustering of metabolic networks.
Each pathway is associated to a function, thus comparing these pathways allows identifying the functional similarities between two networks.
To do so we will compute the intersection of the set of pathways of two (or more) metabolic networks.

Let $M_1$, $M_2$ and $M_{1-2}$ be three metabolic networks such that $M_{1-2}$ is the intersection of $M_1$ and $M_2$. 
Let $P_1$ and $P_2$ be the set of metabolic pathways of $M_1$ and $M_2$. 
Each pathway $p=(V_p,E_p)$ is a subgraph of the network it belongs to.
We denote $name(p)$ the name of $p$.
Then the set $P_{1-2}$ of metabolic pathways of $M_{1-2}$ is defined as follow:
\begin{enumerate}
 \item $\forall p \in P_{1-2}$, $\exists p' \in P_1$ and $p'' \in P_2$ such that $name(p) = name(p') = name(p'')$, and
 \item If $p=(V_p,E_p) \in P_{1-2}$, $p'=(V_{p'},E_{p'})\in P_1$ and $p''=(V_{p''},E_{p''}) \in P_2$ verify the first condition, then $V_p = V_{p'} \cap V_{p''}$ and $E_p=E_{p'} \cap E_{p''}$.
\end{enumerate}

\begin{figure*}
  \centering
  \includegraphics[width=0.7\linewidth]{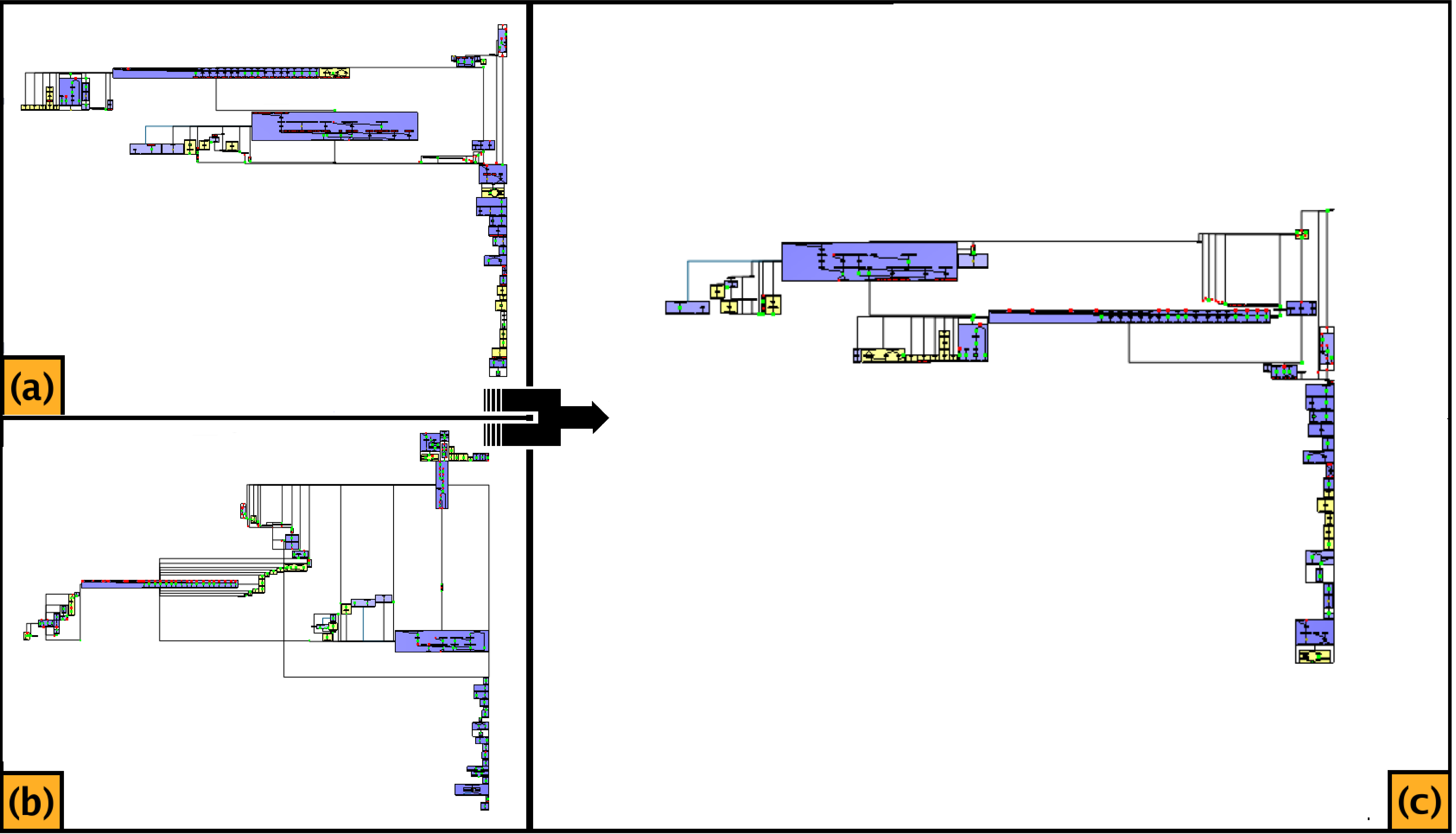}
  \caption{(a) \emph{Rickettsia prowazekii} metabolic network, (b) \emph{Rickettsia typhi} metabolic network and (c) the intersection of (a) and (b) which corresponds to the \emph{typhus} group.}
  \label{meta_networks}
\end{figure*}

This comparison step generates a new network that summarize two or more networks (for instance network (a) and (b) on Figure \ref{meta_networks}).
This simplified view of the metabolism provides a view on core metabolic functions (Figure \ref{meta_networks} (c)).
As our comparison is based on the name of the metabolic pathways, the comparison would be biaised by the different names given to an unique function in several networks. Thus, in the data used in this paper, the pathways of all the networks to compare are named according the same naming rules. 
The question is then to choose which networks we are going to compare.
This will be achieved by using a classification.

\subsection{Building metabolic network hierarchy}
\section{Methodology}
\begin{figure}[ht]
  \centering
  \includegraphics[width=.8 \linewidth]{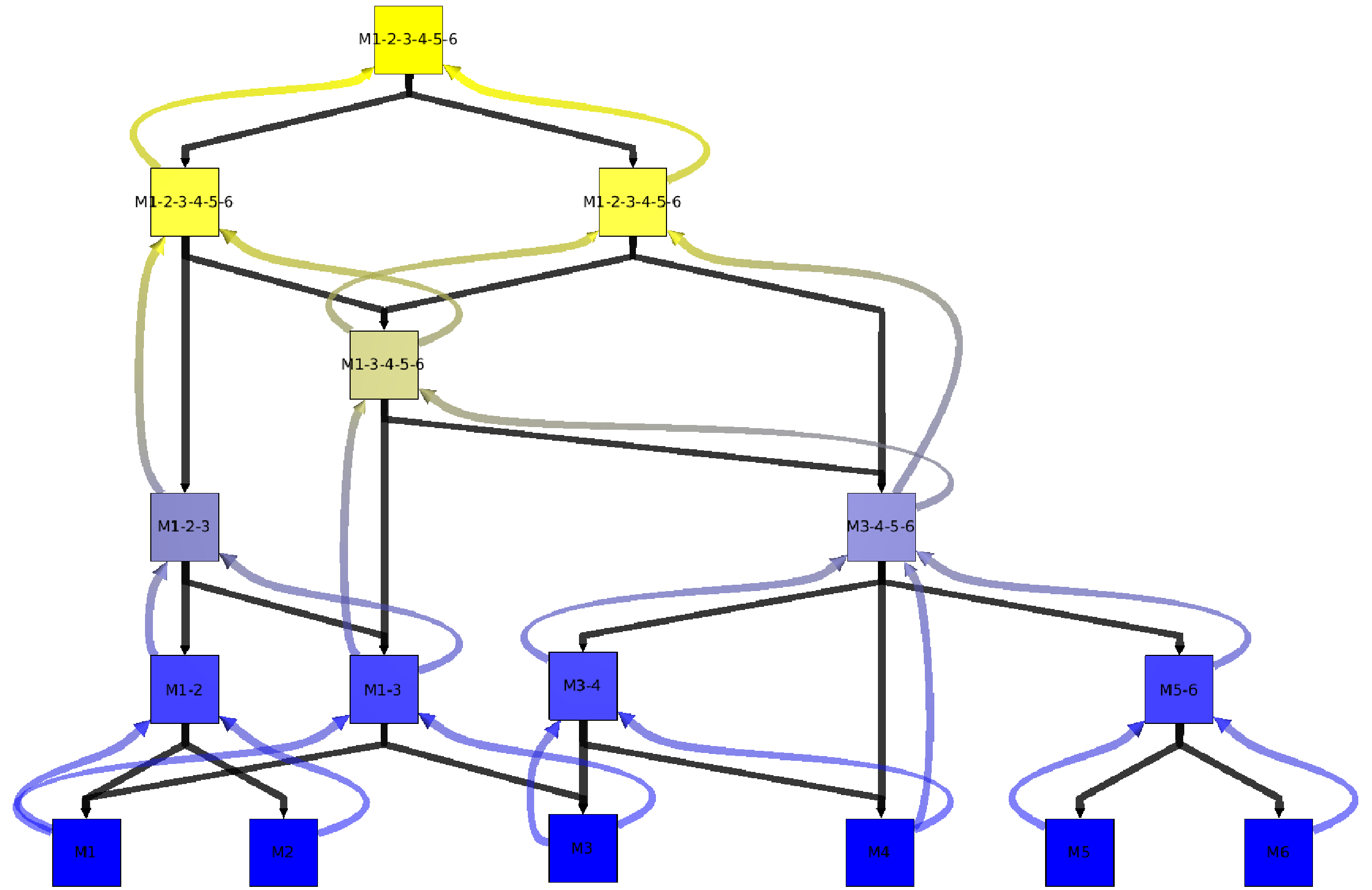}
  \caption{Metabolic network comparisons: the colors show the order in which the metabolic network intersections are computed from dark blue to yellow. Colored arrows indicate which networks are needed to compute the networks of their targets.}
  \label{intersection}
\end{figure}

\begin{figure}[ht]
  \centering
  \includegraphics[width=1. \linewidth]{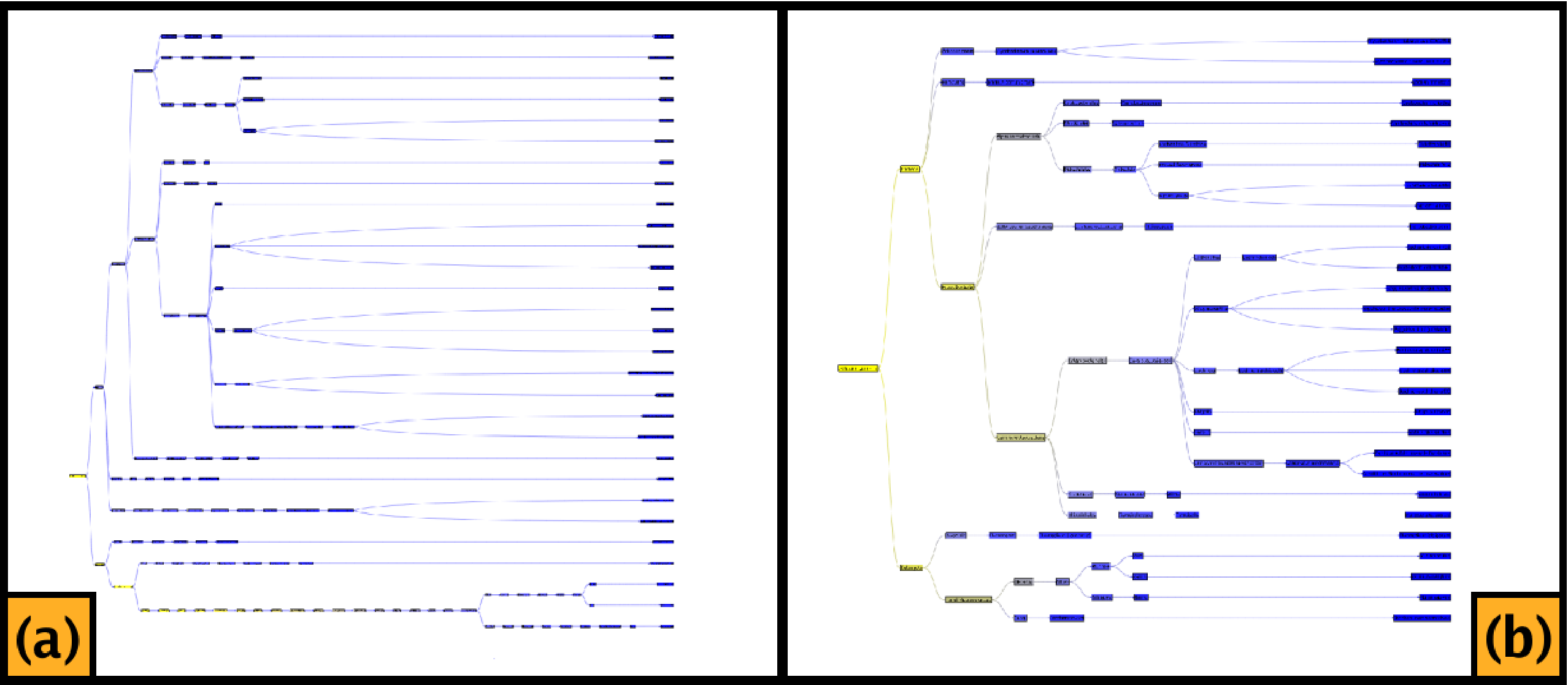}
  \caption{Tree (a) is the taxonomy, described in the NCBI database, for our selected organisms. Tree (b) is the simplified version of the taxonomy.}
  \label{hierarchy_not_simplified}
\end{figure}
To enhance and facilitate the comparison of metabolic networks, we use a hierarchy \emph{DAG} (Directed Acyclic Graph) $H=(V_H,E_H)$ as a visual backbone.
We use a DAG because it is more generic than trees.
In $H$, vertices having an out degree equal to $0$ represent the organisms to compare (\emph{e.g.} leaves in a tree).
Then each network associated to an internal vertex is the result of the comparison of all \emph{underneath} organisms in the hierarchy.
Figure~\ref{intersection} shows in which order metabolic networks are computed (from the firsts in dark blue to the last in yellow) and colored arrows indicate which network are compared.
In a more formal way we define this process as follows.
Let $u$ be a vertex of the $H$ and $N^+(u)$ be the outgoing neighborhood of $u$.
Then the network corresponding to $u$ is the \emph{intersection} of the networks corresponding to all vertices of  $N^+(u)$.
Thus, networks of $N^+(u)$ are needed to compute the network corresponding to $u$. 
We also define $leaves(u)$ as the set of nodes $v$ such that out degree of $v$ is equal to $0$ and there exists a path from $u$ to $v$. Then it is easy to prove that the network corresponding to $u$ is the \emph{intersection} of the metabolic networks corresponding the vertices of $leaves(u)$.

\begin{figure*}
  \centering
  \includegraphics[width=0.8\linewidth]{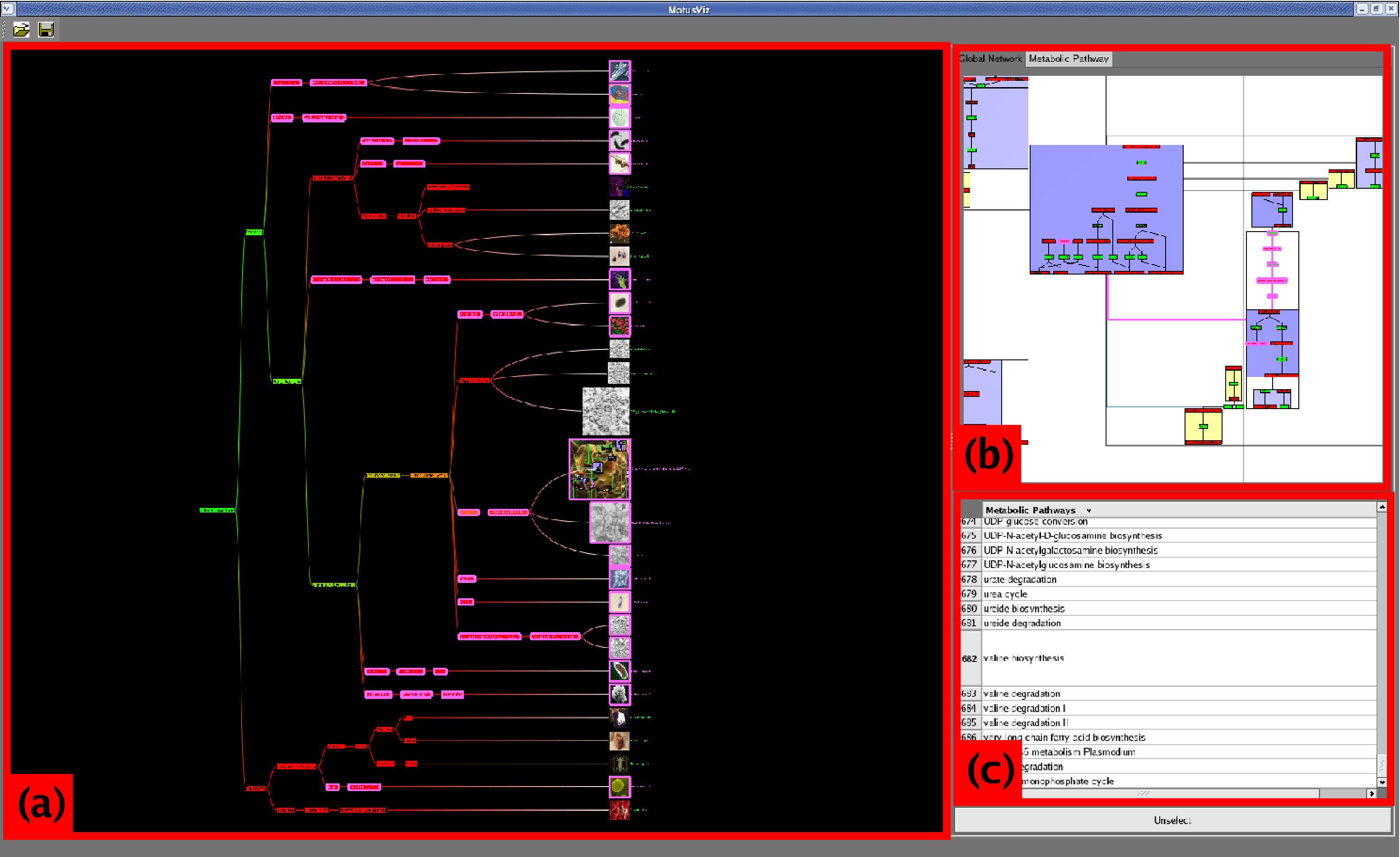}
  \caption{Our visualization tool: (a) View on the hierarchy; (b) Drawing of the focused network, here \emph{Buchnera Aphidicola APS}; (c) List of the names of all metabolic pathways. Here, \emph{Valine biosynthesis} is selected in (c): in hierarchy, networks highlighted in pink contain that pathway. In (b), compounds and reactions of the pathway are highlighted in the focussed network. }
  \label{screenshot}
\end{figure*}

According to the task defined in section~\ref{task}, we chose to use an organism taxonomy tree containing all the organisms to compare.
This hierarchy is coming from the database of the \emph{NCBI} (National Center for Biotechnology Information) database.
The resulting taxonomy tree contains more than $130$ vertices and has a depth equal to $30$.
This tree contains very long branches with no ramification, and thus many vertices don't bring any information (see figure~\ref{hierarchy_not_simplified} (a)).
Since these nodes won't be compared to any other one, we simplify this taxonomy tree by removing them.
To do so, we forbid sequences of vertices $u_1, u_2,..., u_{k-1}, u_k$ of degree equal to $2$ and with $k > 4$ by removing all nodes $u_i$ with $3 \leq i \leq k-2$. Figure~\ref{hierarchy_not_simplified} (b) shows the result of this process: we obtain a simplified taxonomy tree with $81$ vertices and a depth equal to $8$.

\section{Visualization}
As it was mentioned in section \ref{task} the biological task requires visualizing the network and the hierarchy.
To follow biologist representations, mainly inherited from textbooks, we carefully chose our drawing algorithms.
Moreover, due to the large amount of data displayed, we adapted and implemented navigation methods.

\subsection{Drawings}
Representation of a single metabolic network had been intensively investigated in the recent years \cite{M98,BR01,WK05,BLCAMSJ07}.
It is a challenging graph drawing problem for three reasons (upon the ones described in \cite{SND05}). 
Firstly because these networks contain hundred of metabolic pathways made of more than one thousand metabolic reactions.
Secondly it exists drawing constraints (for hierarchy and cycles) defined according to text-book drawings of the pathways.
Finally because biologists expect to be able to visually identify each pathway.
For instance in our visualization, clusters in purple represent metabolic pathways and clusters in yellow represent particular topological structures such as cycles or reaction cascades.
This last point is of utmost importance since biologists aim is to discover a new network according to the pathway it contains. 

To draw the hierarchy, we use two different algorithms, depending on the topology of $H$.
If $H$ is a tree then we use a dendrogram representation of the hierarchy (for instance, see figure~\ref{screenshot}.a).
Otherwise, we use a modified version of the hierarchical  algorithm presented in~\cite{A03}.
This modification consists in laying out all vertices having an out degree equal to $0$ in the same layer, to easily identify the organisms.

The task consists in discovering networks in their context (the hierarchy).
Thus it is necessary to provide a focus plus context facility, that is a way to get both closer views and context representation.
A well suited visualization method is the fisheye method \cite{F81}.
In the next section we describe how we adapted the fisheye to modify smoothly the hierarchical representation.

\subsection{Fisheye for hierarchies}
Fisheye method extends the representation around a focal point and shrink the other part of the view.
It is generally based on a function related to the distance to the focal point \cite{SB94}.
But applying such a method on graph may create edge and/or node overlapping.
To preserve our representation from these artifacts we propose a different way to compute the fisheye view.
This computation is performed in three steps: compute the size of the vertices, apply a drawing algorithm that takes vertex sizes into account and shift the view.

\paragraph{Size of the vertices}
We consider that the focus of the user is at the same position as the mouse pointer.
Thus the size of the vertices is related to their distances to the focus. 
The closer a vertex is to the focus the bigger it is.

\paragraph{Drawing algorithm}
Once the sizes of vertices are modified the hierarchy representation has to be updated.
If it is done without changing the internal node coordinates it may create edge overlapping.
To avoid this problem we compute a new representation of the hierarchy. 
If the hierarchy is a tree, we use a dendrogram algorithm which needs to go through all the $m$ edges and $n$ nodes since it takes into account their size.
Therefore each update of the display can be done in $O(m+n)$.

If the hierarchy is a DAG and if we directly use a classical DAG drawing algorithm we can reach a complexity of $O(m\times n)$.
This complexity is too high to get a smooth display.
But taking into account the specificity of our visualization we are going to see that it can be computed it in $O(m+n)$.
We base our method on the hierarchical algorithm presented in~\cite{A03} which works as follow: first, vertices are assigned to layers, then vertices of each layer are ordered to minimize the number of edge crossings, and finally coordinates are assigned to each vertex. 
To update the displaying, we just need to recompute the last step of this algorithm since layers assigned to the vertices and the order in each layer do not change. Trivially layer placement and node placement in each layer can be done in $O(m+n)$.

\paragraph{Layout shifting}
As vertex sizes are modified, all the vertices are getting further from their original position, thus affecting user mental map.
Moreover these coordinate modifications can be increased by the number of focal point changes.
To bound the number of moves in the drawing, we constraint vertice positions. This is done by translating the hierarchy such that the focus vertex position is set to its original position.
And then, before updating the display, the view is shifted such that the distance between the mouse pointer and the focus is kept unchanged.
Consequently, the user mental map is preserved.

\subsection{Navigation through the different scales}

Figure~\ref{screenshot} shows a screenshot of our visualization tool. This tool contains three main widgets. First, a view on the hierarchy is shown on figure~\ref{screenshot}.a. Then in the top right corner (see figure~\ref{screenshot}.b), we can see a view on a metabolic network. And finally, in the bottom right corner there is a list of all pathways contained in the organisms to compare. There also exists a view at metabolic pathway level since widget (b) of figure~\ref{screenshot} allows switching from a metabolic network view to a metabolic pathway view.

Therefore, our visualization tool allows navigating from the highest level (the hierarchy) to the lowest level (the metabolic pathways).

\paragraph{Highest level: Hierarchy}
All pathways of all organisms to compare are listed in widget (c) of figure~\ref{screenshot}. Clicking on one of these pathways allows focusing on a given pathway. Then, in order to know which networks of the hierarchy contain the current pathway, they are highlighted in pink. For instance, on figure~\ref{screenshot}, we selected the \emph{Valine Biosynthesis} pathway, we can see that the synthesis of this essential amino-acid is not present in all organisms: for instance no \emph{Rickettsiales} can synthesize it. On the contrary, \emph{Caulobacter vibroides} and \emph{Agrobacterium tumefaciens}, i.e. the others \emph{Alphaproteobacteria} of the set of organisms, can synthesize it.
It is thus possible to discriminate \emph{Rickettsiales} from other organisms or, using the hierarchy, class of organisms.
\begin{figure}[h!]
  \centering
  \includegraphics[width=.8\linewidth]{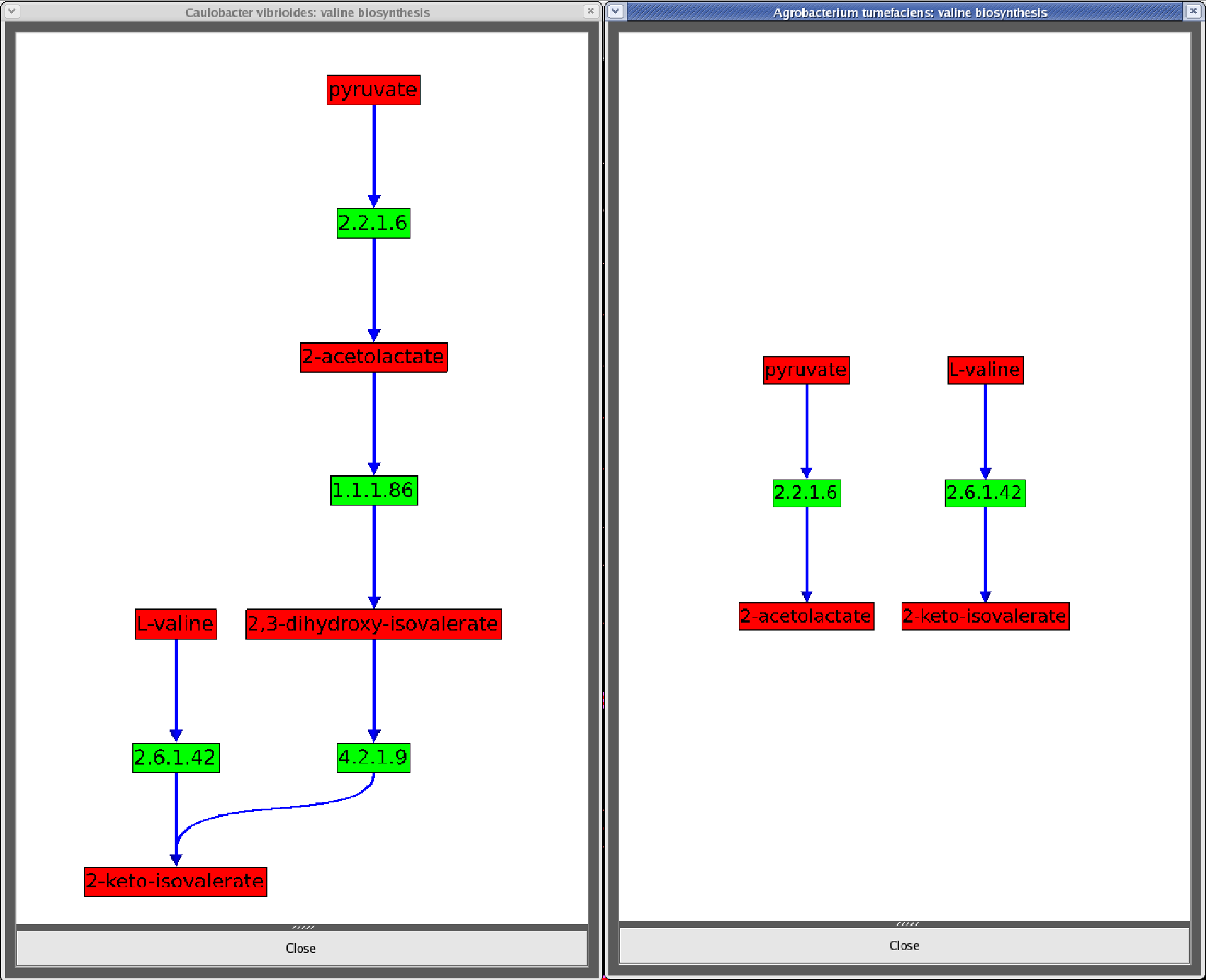}
  \caption{Focus on \emph{Valine biosynthesis} pathway: on the left the corresponding pathway in \emph{Caulobacter vibroides}, and on the right those of \emph{Agrobacterium tumefaciens}.}
  \label{case_study_2}
\end{figure}

\paragraph{Intermediate level: Metabolic network}\label{met_level}
To focus on a particular metabolic network, the user can use a \emph{fisheyes} distortion on the hierarchy (in figure~\ref{screenshot}, organism in the center of the fisheye is \emph{Buchnera Aphidicola APS}). The vertex $u$ laid out the closest to the center of the fisheye is focused. Then the metabolic network represented by $u$ is displayed in widget (b) in figure~\ref{screenshot}).

If a metabolic pathway had been selected and if the focused organism contains this pathway, then it is highlighted in pink in the metabolic network view. It allows visualizing a pathway in its context (in figure~\ref{screenshot}, the \emph{Valine Biosynthesis} pathway is highlighted in \emph{Buchnera Aphidicola}).

\paragraph{Lowest level: Metabolic pathway}
At the lowest level, the user can compare metabolic pathways in different organism.
We propose this facility since two pathways having the same name may not contain exactly the same compounds and reactions.
Indeed there can be several ways to product the same compound.
When a pathway is selected, \emph{double-clicking} on an organism containing this pathway allows to visualize the corresponding pathway of this organism. Figure~\ref{case_study_2} shows metabolic pathways \emph{Valine Biosynthesis} in \emph{Caulobacter vibroides} and \emph{Agrobacterium tumefaciens}. We can see that these two pathways share two reactions but also that two of them are missing in the pathway of \emph{Agrobacterium tumefaciens} (the reactions $1.1.1.86$ and $4.2.1.9$).

\section{Conclusion}
In this article we propose a generic way to take advantage of network clustering when visually comparing networks.
Moreover we show that adding a context to the visualization helps in understanding the data.
In fact, the context helps to choose the comparison to perform.
Moreover it provides clues on the biological families and how they share metabolic pathways.
We present an implementation of this method for a particular biology study case.
Based on this representation, biologists were able to put new biological networks into a wider context.

Since this project was done in collaboration with biologists focused on specific networks we were not able to test our approach on other data.
We plan for instance to use it on protein-protein interaction networks. We also plan to adapt this method to other domains such as indexed videos (or images) collection to facilitate the research of a given video (or images).
On a visualization point of view we are improving the navigation by implementing an algorithm of pathways alignment \cite{BDS04}, it will highlight the common reactions and compounds of several pathways.

\section*{Acknowledgments}
Ludovic Cottret for his advises on the biological case study.
Project funded by the ANR-BBSR project Systryp.
\bibliographystyle{latex8}
\bibliography{iv08bib}

\begin{thebibliography}{10}\setlength{\itemsep}{-1ex}\small

\bibitem{A03}
D.~Auber.
\newblock Tulip : {A} huge graph visualisation framework.
\newblock In {\em Graph Drawing Softwares}, Mathematics and Visualization,
  pages 105--126. Springer-Verlag, 2003.

\bibitem{BR01}
M.~Becker and I.~Rojas.
\newblock A {G}raph {L}ayout {A}lgorithm for {D}rawing {M}etabolic {P}athways.
\newblock {\em Bioinformatics}, 17:461--467, 2001.

\bibitem{BLCAMSJ07}
R.~Bourqui, V.~Lacroix, L.~Cottret, D.~Auber, P.~Mary, M.-F. Sagot, and
  F.~Jourdan.
\newblock Metabolic network visualization eliminating node redundance and
  preserving metabolic pathways.
\newblock {\em BMC Systems Biology}, 1(29), 2007.

\bibitem{BDS04}
U.~Brandes, T.~Dwyer, and F.~Schreiber.
\newblock Visualizing {R}elated {M}etabolic {P}athways in {T}wo and {H}alf
  {D}imensions.
\newblock {\em LNCS}, 2912:11--122, 2004.

\bibitem{F81}
G.~W. Furnas.
\newblock The {FISHEYE} view: {A} new look at structured files.
\newblock Technical report, Murray Hill, U.S.A., 1981.

\bibitem{GJ79}
M.~R. Garey and D.~S. Johnson.
\newblock {\em Computers and {I}ntractability: {A} {G}uide to the {T}heory of
  {NP}-{C}ompleteness}.
\newblock W. H. Freeman \& Co., New York, NY, USA, 1979.

\bibitem{Jourdan2008}
F.~Jourdan, R.~Breitling, M.~Barrett, and D.~Gilbert.
\newblock Meta{N}etter: inference and visualization of high-resolution
  metabolomic networks.
\newblock {\em Bioinformatics}, 24:143--145, 2008.

\bibitem{biocyc}
P.~Karp, C.~Ouzounis, C.~Moore-Kochlacs, L.~Goldovsky, P.~Kaipa, D.~Ahren,
  S.~Tsoka, N.~Darzentas, V.~Kunin, and N.~Lopez-Bigas.
\newblock Expansion of the biocyc collection of pathway/genome databases to 160
  genomes.
\newblock {\em Nucleic Acids Research}, 19:6083--6089, 2005.

\bibitem{M98}
G.~Michal.
\newblock On representation of metabolic pathways.
\newblock {\em BioSystems}, 47:1--7, 1998.

\bibitem{Pinter05}
R.~Pinter, O.~Rokhlenko, E.~Y.-L. E, and M.~Ziv-Ukelson.
\newblock Alignment of metabolic pathways.
\newblock {\em Bioinformatics}, 21:3401–3408, 2005.

\bibitem{SND05}
P.~Saraiya, C.~North, and K.~Duca.
\newblock Visualizing biological pathways: requirements analysis, systems
  evaluation and research agenda.
\newblock {\em Information Visualization}, pages 1--15, 2005.

\bibitem{SB94}
M.~Sarkar and M.~H. Brown.
\newblock Graphical fisheye views.
\newblock {\em Communications of the ACM}, 37(12):73--84, 1994.

\bibitem{VWGW02}
J.~van Helden, L.~Wernisch, D.~Gilbert, and S.~Wodak.
\newblock Graph-based analysis of metabolic networks.
\newblock {\em Ernst Schering Research Foundation Workshop}, 38:245--274, 2002.

\bibitem{WK05}
K.~Wegner and U.~Kummer.
\newblock A new dynamical layout algoritmh for complex biochemical reaction
  networks.
\newblock {\em BMC Bioinformatics}, 2005.

\end{thebibliography}

\end{document}